\documentclass[11pt,a4paper]{article}
\usepackage[T1]{fontenc}
\usepackage[utf8]{inputenc}
\usepackage{lmodern}
\usepackage[margin=2.5cm]{geometry}
\usepackage{amsmath,amssymb}
\usepackage{graphicx}
\usepackage{booktabs}
\usepackage{tabularx}
\usepackage{array}
\usepackage{float}
\usepackage{siunitx}
\usepackage{listings}
\usepackage{pdflscape}
\usepackage{xcolor}
\usepackage{colortbl}
\usepackage{microtype}
\usepackage{textcomp}
\usepackage[numbers]{natbib}
\usepackage{hyperref}
\usepackage{xurl}
\usepackage{cleveref}

\newcolumntype{L}[1]{>{\raggedright\arraybackslash}p{#1}}
\newcolumntype{C}{>{\centering\arraybackslash}X}
\newcommand{\repolink}[2]{\href{#1}{\footnotesize #2}}
\newcommand{\red}[1]{}

\newcommand{\authorcontributions}[1]{\section*{Author Contributions}#1}
\newcommand{\funding}[1]{\section*{Funding}#1}
\newcommand{\institutionalreview}[1]{\section*{Institutional Review Board Statement}#1}
\newcommand{\informedconsent}[1]{\section*{Informed Consent Statement}#1}
\newcommand{\dataavailability}[1]{\section*{Data Availability Statement}#1}
\newcommand{\acknowledgments}[1]{\section*{Acknowledgments}#1}
\newcommand{\conflictsofinterest}[1]{\section*{Conflicts of Interest}#1}

\newcommand{\abbreviations}[2]{\section*{#1}#2}

\definecolor{codeblue}{rgb}{0.05,0.10,0.65}
\definecolor{codeteal}{rgb}{0.00,0.42,0.45}
\definecolor{codepurple}{rgb}{0.50,0.00,0.50}
\definecolor{codegreen}{rgb}{0.00,0.45,0.00}
\definecolor{codegray}{rgb}{0.45,0.45,0.45}
\definecolor{codestring}{rgb}{0.65,0.15,0.00}

\lstdefinelanguage{Rust}{
  sensitive=true,
  keywords={as, async, await, break, const, continue, crate, dyn, else, enum,
    extern, false, fn, for, if, impl, in, let, loop, match, mod, move, mut,
    pub, ref, return, self, Self, static, struct, super, trait, true, type,
    unsafe, use, where, while},
  keywords=[2]{bool, char, f32, f64, i8, i16, i32, i64, i128, isize, str,
    u8, u16, u32, u64, u128, usize, String, Vec, Option, Result, Some, None,
    Ok, Err, Box},
  keywords=[3]{array},
  keywordstyle=\color{codeblue}\bfseries,
  keywordstyle=[2]\color{codeteal},
  keywordstyle=[3]\color{codepurple},
  comment=[l]{//},
  morecomment=[s]{/*}{*/},
  commentstyle=\color{codegray}\itshape,
  stringstyle=\color{codestring},
  morestring=[b]",
}

\newcommand{\matchedPairs}{1320}
\newcommand{\fasterPct}{99.8}
\newcommand{\geomeanSpeedup}{9.3}
\newcommand{\geomeanSpeedupPm}{0.9}
\newcommand{\geomeanVsPython}{0.107}
\newcommand{\geomeanVsPythonPm}{0.010}
\newcommand{\gpuSpeedupMin}{33.5}
\newcommand{\gpuSpeedupMax}{200}
\newcommand{\sigTotal}{275}
\newcommand{\sigFaster}{272}
\newcommand{\sigPar}{3}
\newcommand{\sigSlower}{0}
\newcommand{\sigFasterPct}{98.9}
\newcommand{\sigParPct}{1.1}

\newcommand{\exKuramotoSeconds}{22.14}
\newcommand{\exKuramotoCalls}{75}
\newcommand{\exLogisticSeconds}{1.27}
\newcommand{\exLogisticCalls}{753}
\newcommand{\exRosslerEpsSeconds}{74.08}
\newcommand{\exRosslerEpsCalls}{1260}
\newcommand{\exRosslerRenyiSeconds}{0.11}
\newcommand{\exRosslerRenyiCalls}{30}
\newcommand{\exRosslerTsallisSeconds}{0.11}
\newcommand{\exRosslerTsallisCalls}{30}

\newcommand{\lstEntropy}{0.6931}
\newcommand{\lstMi}{0.5095}
\newcommand{\lstTe}{0.3213}

\title{\textit{infomeasure-rs}: A High-Performance Rust Library for
Information-Theoretic Measures and Estimators}
\author{Carlson Moses B\"uth$^{1,2}$\thanks{ORCID: 0000-0003-2298-8438.}\\
\small $^{1}$ Institute for Cross-Disciplinary Physics and Complex Systems (IFISC), CSIC-UIB,\\
\small \phantom{$^{1}$}Palma de Mallorca, Spain\\
\small $^{2}$ Department of Computational Linguistics, University of Zurich,\\
\small \phantom{$^{2}$}Andreasstrasse 15, 8050 Zurich, Switzerland}
\date{23 November 2026}

\begin{document}
\maketitle

\begin{abstract}
Information theory has become a standard tool across the natural and
social sciences, yet practical use is limited by estimator choice, fragmented
software, reproducibility, and computational cost. We
present \textit{infomeasure-rs}, an open-source Rust library that brings a
broad set of information-theoretic measures and estimators into one
self-contained, type-safe package.
It covers entropy, mutual information, conditional mutual information, transfer entropy, conditional transfer entropy,
cross-entropy, and the Kullback--Leibler and Jensen--Shannon divergences, with
discrete (plug-in and bias-corrected), kernel, ordinal, and nearest-neighbour
estimators, including the R\'enyi and Tsallis entropies, and exposes local and
global values and significance testing through a consistent trait interface.
It compiles to self-contained native code with no language runtime dependency and is continuously validated against
a reference implementation. On a
single central processing unit (CPU) core, and even without its GPU or parallel
paths, it is the fastest of the packages we compared across languages,
matching the closest packages only for plug-in discrete entropy and
substantially faster for mutual information, the conditional and transfer
measures, and the kernel, ordinal, and generalised-entropy estimators. The optional GPU and
multi-threaded paths accelerate selected estimators further. The library is open source and, to our knowledge, the most complete
information-theory package for Rust.
\end{abstract}


\begin{center}
\textit{Accepted for publication in \emph{Entropy} (Special Issue: ``Entropy-Based Time Series Analysis: Theory and Applications'').}
\end{center}

\section{Introduction}

Information theory, the~mathematical study of information and its processing~\cite{shannon1948mathematical}, has become a cornerstone of modern data
analysis, with~applications ranging from neuroscience~\cite{vicente2011transfer,wibral2014directed} and complex systems~\cite{zanin2017network} to finance~\cite{zhou2013applications}. Applying it in
practice, however, is rarely as simple as choosing a measure and computing it.
Four barriers recur~\cite{buthInfomeasureComprehensivePython2025a}. First,
information-theoretic quantities are functionals of probability
distributions that are unknown in practice and must be estimated from finite
samples. The~choice of estimator strongly affects the result, and~limited
guidelines exist for selecting it~\cite{de2024entropy,contreras2021selecting}. Second, the~software landscape is
fragmented: individual packages tend to concentrate on a subset of measures or
estimators, which makes systematic comparison cumbersome. Third,
implementation-level differences between packages hinder reproducibility.
Fourth, the~estimators are computationally intensive, and~thus large data sets are
prohibitive unless the implementation is carefully~optimised.

The Python package \textit{infomeasure} (version 0.6.3) was introduced to lower these barriers
by unifying a broad set of measures and estimators under one consistent
interface~\cite{buthInfomeasureComprehensivePython2025a}. A~year after its
introduction, it is used in research across several fields, including neuroscience~\cite{tiawongsuwan2025autism,garcialeal2026functional,martinezvasquez2026frontal,goffinet2026torus},
transport-delay analysis~\cite{buth2025-synthatdelays,buthFunctionalConnectivityNetworks2025}, energy and
industrial monitoring~\cite{belhadj2026inferring,luo2026hierarchical}, biology
and biophysics~\cite{iakhiaev2026tissue,beker2026morphogenesis}, security and
privacy~\cite{berke2026fingerprinting}, the~study of complex systems
~\cite{zanin2025limits}, and~methodological work on information estimation
~\cite{mccabe2026seneca,bulinski2026statistical,abbasi2026eventropy,bayesianica2026,alternatingdiffusion2026}.
This highlights the importance that such unifying implementations have in the
scientific community, as~they lower the implementation cost and enhance
reproducibility. For~the most demanding workloads, the~Python runtime is the
limiting factor, and~a compiled implementation removes it while keeping the
numerical results identical. The~same compiled library can serve as a backend for the Python package, so existing
users gain the speed without changing their~workflow.

We present \textit{infomeasure-rs}, a~Rust library that reimplements the
\textit{infomeasure} measure and estimator set from first principles. Rust
offers the performance of a compiled language together with memory safety and
predictable semantics. The~library compiles to self-contained native code and
imposes no language runtime or interpreter. The~Rust information-theory
ecosystem is still young, and~we survey it in Section~\ref{sec:related}.

The contribution is a complete, validated, high-performance implementation for
the Rust ecosystem, together with the benchmark and validation infrastructure
that makes it comparable with established packages. Specifically, we provide the~following:
\begin{itemize}
\item A complete Rust library covering the measures and estimators of the
      Python reference, exposed through a type-safe trait interface.
\item A continuous validation strategy that re-checks every estimator against
      the validated Python implementation on each change.
\item A fair, reproducible cross-package benchmark (identical inputs, identical
      parameters, a~single CPU thread).
\item An average speed-up of about ten times over the Python implementation on
      a single thread, with~optional GPU and multi-threaded acceleration.
\end{itemize}

\section{Existing~Software}
\label{sec:related}

The companion Python paper~\cite{buthInfomeasureComprehensivePython2025a}
already places \textit{infomeasure} among the established Python, Java, Julia
and R toolkits, including the Java Information Dynamics Toolkit (JIDT)~\cite{lizierJIDTInformationTheoreticToolkit2014},
PyInform~\cite{mooreELIFEASUPyInformV0202019}, pyitlib~\cite{fosterPyitlibLibraryInformationtheoretic}, dit~\cite{jamesDitPythonPackage2018}, Syntropy~\cite{varley2025syntropy},
DiscreteEntropy.jl~\cite{kellyDiscreteEntropyjlEntropyEstimation2024}, pyEntrp~\cite{donetsNikdonPyEntropy2025}, and~RTransferEntropy~\cite{behrendt2019rtransfer}. We refer the reader there for that comparison
and do not repeat it here. This note instead addresses the ecosystem that its
title concerns: the state of information-theoretic software in Rust.

\label{sec:rust-survey}

To situate \textit{infomeasure-rs} honestly we surveyed crates.io,
Rusts central package registry,
on 1 October 2026 using the search terms \textit{entropy}, 
 \textit{Shannon entropy},
\textit{R\'enyi entropy}, \textit{Tsallis entropy}, \textit{mutual
information}, \textit{transfer entropy}, and~\textit{information theory}, then
inspected every plausible hit. Because~the search is keyword based it is a
bounded observation rather than an exhaustive census, but~the picture is
clear and is summarised in Table~\ref{tab:rustsurvey}.

\clearpage
\begin{landscape} 
\begin{table}[H]
\centering
\caption{Representative
 information-theoretic crates on crates.io, from a
keyword survey on 2026-10-01. ``Estimators'' describes what the crate actually
implements, and it is not a completeness claim. ``Updated'' is the month of the most
recent release on crates.io. The project described in this note is
\textit{infomeasure-rs}, distributed as the crate \texttt{infomeasure}.
\label{tab:rustsurvey}}
\small
\begin{tabularx}{\linewidth}{L{4cm} L{4.5cm} X L{7.3cm} c}
\toprule
\textbf{Crate} & \textbf{Measures} & \textbf{Estimators/Scope} & \textbf{Link} & \textbf{Updated}\\
\midrule
\texttt{infomeasure} & H, MI, CMI, TE, CTE, KLD, JSD & discrete (plug-in and bias-corrected), kernel, ordinal, kNN/KSG, R\'enyi, Tsallis & \repolink{https://crates.io/crates/infomeasure}{crates.io/crates/infomeasure} & 2026-10 \\
\texttt{rstats} & general statistics & some information measures, broad scope & \repolink{https://github.com/liborty/rstats}{github.com/liborty/rstats} & 2026-08 \\
\texttt{logp} & H, MI, KLD, divergences & KSG for MI, plug-in otherwise & \repolink{https://github.com/arclabs561/logp}{github.com/arclabs561/logp} & 2026-06 \\
\texttt{lau-information-theory} & H, MI, channel capacity & discrete plug-in & \repolink{https://github.com/SuperInstance/lau-information-theory}{github.com/SuperInstance/lau-information-theory} & 2026-06 \\
\texttt{channel-capacity} & MI, channel capacity & discrete memoryless channels & \repolink{https://github.com/SuperInstance/channel-capacity}{github.com/SuperInstance/channel-capacity} & 2026-06 \\
\texttt{huffman-entropy} & H, R\'enyi, Tsallis & distribution-level, coding oriented & \repolink{https://github.com/SuperInstance/huffman-entropy}{github.com/SuperInstance/huffman-entropy} & 2026-06 \\
\texttt{entropium} & H, joint H, conditional H, MI & plug-in over distributions & \repolink{https://crates.io/crates/entropium}{crates.io/crates/entropium} & 2026-04 \\
\texttt{infotheory} & algorithmic information & Kolmogorov complexity, not estimators & \repolink{https://github.com/turtle261/infotheory}{github.com/turtle261/infotheory} & 2026-03 \\
\texttt{entropy} & Shannon H & bytes and strings, no estimator & \repolink{https://github.com/smackysnacks/entropy}{github.com/smackysnacks/entropy} & 2026-01 \\
\texttt{nmi} & normalised MI & clustering score & \repolink{https://github.com/jean-pierreBoth/coreset}{github.com/jean-pierreBoth/coreset} & 2025-06 \\
\texttt{information} & H, MI and related & functions over \texttt{ndarray} & \repolink{https://github.com/noamteyssier/information}{github.com/noamteyssier/information} & 2022-10 \\
\texttt{entropy-lib} & Shannon / metric H & files, no estimator & \repolink{https://github.com/tuxxy/entropy}{github.com/tuxxy/entropy} & 2020-05 \\
\texttt{shannon-entropy} & Shannon H & strings, no estimator & \repolink{https://github.com/insanitybit/shannon-entropy}{github.com/insanitybit/shannon-entropy} & 2020-03 \\
\bottomrule
\end{tabularx}
\end{table}
\end{landscape}

The Rust ecosystem is young and fragmented. The~most downloaded crates compute
the Shannon entropy of byte strings or files, a~task for which no estimator is
involved. A~second group wraps a single estimator of a single measure, most
often entropy or (as a feature score) mutual information. Most of the crates we
found are recent, but~each covers only a narrow slice, and~a few have not been
updated in years. A~handful provide information-theory primitives, and~one
(\texttt{logp}) includes a KSG mutual-information estimator, but~none of those
we found provides conditional mutual information, transfer entropy, or~conditional transfer entropy, and~none offers the range of estimator
families that the Python reference exposes: bias-corrected discrete, kernel,
ordinal, and~nearest-neighbour with R\'enyi and Tsallis variants.
General-purpose statistics and dynamics crates expose isolated entropy
routines but not the estimator interface that makes the measures selectable and
comparable. In~short, the~Rust language has no equivalent of the multi-measure,
multi-estimator toolkits that exist in the other ecosystems.
\textit{infomeasure-rs} is intended to be that~equivalent.

Of these, only \texttt{logp} and \texttt{entropium} implement measures that are
directly comparable to ours (a KSG mutual-information estimator and plug-in
discrete entropies, respectively) and are candidates for the runtime comparison
of Section~\ref{sec:perf}. The~remainder are byte-string entropy utilities or
single-purpose or stale crates, or~address a different problem (algorithmic
complexity, channel capacity, clustering scores, or~a broad general-purpose
statistics library), and~fall outside the benchmark's scope for the same
reasons the Python survey excludes analogous~packages.

Underpinning the nearest-neighbour searches is \texttt{kiddo}, a~k-d tree
library~\cite{kiddo}. C.M.B. contributed to its development: distance metrics and
neighbour-query boundary handling in v5.3.0, and~small-$k$ result-collection and
query-path performance work in the v6 series (v6.0.0-alpha.2). The~latter were
motivated by profiling the v6 migration, which initially regressed several
estimators. These primitives are used by the estimators here, but~the speed-up
is not attributable to a single component: it results from the library's
estimator-level algorithms and data structures and their integration with
\texttt{kiddo}. The~library is also fastest for the non-$k$-NN families, which
do not use k-d trees. For~the expensive Gaussian kernel, this
co-development lets \textit{infomeasure-rs} match or exceed the Python reference
even though the latter evaluates the density through compiled Fortran~routines.

\section{The \textit{infomeasure-rs} Package}
\label{sec:package}

The library is distributed on crates.io and documented on docs.rs, following
the open-source practice of the companion Python package~\cite{buthInfomeasureComprehensivePython2025a}. The~crates.io package and Rust
crate are named \texttt{infomeasure}, while the project and repository are named
\textit{infomeasure-rs}. The~practical links, documentation, and~community
guidelines are collected at the end of this section (Section~\ref{sec:meta}).

\subsection{Measures and~Estimators}

The library computes entropy, joint and cross-entropy, mutual
information (MI), conditional mutual information (CMI), transfer entropy (TE),
conditional transfer entropy (CTE), and~the Kullback--Leibler and
Jensen--Shannon divergences~\cite{shannon1948mathematical,kullback1951information,lin1991divergence}. The~divergences are first-class composite measures, gated by traits rather than a
runtime whitelist: KLD is provided through an extension trait for every
estimator that supports cross-entropy, and~JSD, which needs a well-defined
mixture, through a probability-mass capability trait for the estimators that
expose a normalised mass function, with~a pooled kernel density in the
continuous case. The~trait bounds are designed so that combinations without a
sound definition are rejected at compile time.
Table~\ref{tab:coverage} summarises the estimator coverage. R\'enyi and Tsallis
generalizations~\cite{renyi1961measures,tsallis1988possible} are available
through the nearest-neighbour family~\cite{leonenkoClassRenyiInformation2008},
and the bias-corrected discrete estimators follow the overview in
Refs.~\cite{de2024entropy,contreras2021selecting}.

\begin{table}[H]
\caption{Measures and estimation approaches in \textit{infomeasure-rs}, mirroring
the Python package. A checkmark indicates that the estimator is available for the
measure. Estimators marked with a dagger are available for a subset of
bias-corrected discrete variants. \label{tab:coverage}}
\footnotesize
\setlength{\tabcolsep}{2pt}
\begin{tabularx}{\textwidth}{@{}l l C C C C C C C@{}}
\toprule
\textbf{Measure} & \textbf{Notation} & {\scriptsize\textbf{Discrete}} & {\scriptsize\textbf{Kernel}} & {\scriptsize\textbf{kNN/KSG}} & {\scriptsize\textbf{Ordinal}} & {\scriptsize\textbf{Bias-corr.}} & {\scriptsize\textbf{R\'enyi}~$\ast$} & {\scriptsize\textbf{Tsallis}~$\ast$}\\
\midrule
Entropy & $H(X)$ & \checkmark & \checkmark & \checkmark & \checkmark & \checkmark & \checkmark & \checkmark\\
Joint/cross-entropy & $H(X,Y)$/$H_Q(P)$ & \checkmark & \checkmark & \checkmark & \checkmark & \checkmark$\dagger$ & \checkmark & \checkmark\\
Mutual information & $I(X;Y)$ & \checkmark & \checkmark & \checkmark & \checkmark & \checkmark & \checkmark & \checkmark\\
Conditional MI & $I(X;Y \mid Z)$ & \checkmark & \checkmark & \checkmark & \checkmark & \checkmark & \checkmark & \checkmark\\
Transfer entropy & $T_{X \to Y}$ & \checkmark & \checkmark & \checkmark & \checkmark & \checkmark & \checkmark & \checkmark\\
Conditional TE & $T_{X \to Y \mid Z}$ & \checkmark & \checkmark & \checkmark & \checkmark & \checkmark & \checkmark & \checkmark\\
KLD & $D_{\mathrm{KL}}(P \parallel Q)$ & \checkmark & \checkmark & \checkmark & \checkmark & \checkmark$\dagger$ & \checkmark & \checkmark\\
JSD & $\mathrm{JSD}(P \parallel Q)$ & \checkmark & \checkmark & & \checkmark & \checkmark$\ddagger$ & & \\
\bottomrule
\end{tabularx}
\noindent{\footnotesize{$\ast$~The R\'enyi and Tsallis columns denote the
parametric generalizations, evaluated with the nearest-neighbour estimators.
$\dagger$~KLD additionally covers the Miller--Madow and Bayes discrete
variants. $\ddagger$~JSD covers the Bayes and shrinkage discrete variants. JSD
requires a well-defined mixture, so it is not available for the
differential/generalised estimators (Kozachenko--Leonenko, R\'enyi, Tsallis).}}
\end{table}

The discrete family spans the plug-in (maximum-likelihood) estimator and the
bias-corrected Miller--Madow~\cite{miller1955note}, Grassberger~\cite{grassbergerFiniteSampleCorrections1988,grassbergerEntropyEstimatesInsufficient2008}
and James--Stein shrinkage~\cite{hausserEntropyInferenceJamesStein2009}
estimators; the coverage-based Chao--Shen~\cite{chaoNonparametricEstimationShannons2003} and Chao--Wang--Jost~\cite{chaoEntropySpeciesAccumulation2013} estimators; the Bayesian Bayes~\cite{bayesEssaySolvingProblem1763} and Nemenman--Shafee--Bialek (NSB)
\cite{nemenmanEntropyInferenceRevisited2002} estimators, together with the
asymptotic NSB (ANSB) \cite{nemenmanEntropyInformationNeural2004}; and the
specialized Bonachela~\cite{bonachelaEntropyEstimatesSmall2008} and Zhang
~\cite{grabchakAuthorshipAttributionUsing2013,lozanoFastCalculationEntropy2017}estimators. The~kernel family supports box and Gaussian kernels
~\cite{silverman1986density}. The~nearest-neighbour family implements theKozachenko--Leonenko entropy~\cite{kozachenko1987sample} and the
Kraskov--St\"ogbauer--Grassberger mutual information~\cite{miKSG2004}, and~the ordinal family implements permutation-based
estimation~\cite{bandt2002permutation}. TE and CTE use the standard
delay-embedding construction~\cite{schreiber2000measuring}. The~measure
definitions and the limits of the generalised families are collected in
Appendix~\ref{app:definitions}.

\subsection{Statistical Significance~Testing}

An estimate alone does not indicate whether a dependence is real. The~library
therefore provides significance tests for the mutual-information and
transfer-entropy measures, including their conditional variants, across every
estimator family. Let $\mathcal{T}(X,Y,Z)$ be the estimated measure for source
$X$, target $Y$, and~conditioning variables $Z$, and~let\linebreak  
$t=\mathcal{T}(X,Y,Z)$ be the observed value. The~permutation test asks whether
$X$ is independent of $Y$ given $Z$. The~null distribution is built from
$\mathcal{T}(X^{\pi},Y,Z)$ over resamples $\pi$, and~the $p$-value is the
fraction of null values strictly greater than $t$. A~permutation shuffle preserves each marginal distribution but destroys dependence,
so it targets independent and identically distributed (iid) exchangeability
~\cite{schreiber2000measuring}. The~current
bootstrap resamples the source with replacement while holding the target fixed.
This breaks the source-target pairing, so it is a null-like resampling rather
than a joint bootstrap, and~it should not be read as a confidence interval for
the estimate. Both are available and the permutation variant is the default. For~autocorrelated series we additionally provide block permutation, circular shifts,
and a block bootstrap, which preserve short-range source structure. The~block
bootstrap is also a source-resampling null, not a joint bootstrap. The~block
length should reflect the autocorrelation time. The~API was introduced in
version~0.6.0. Each test returns the $p$-value, the~$t$-value or standardised
effect size, the~mean and standard deviation of the null distribution and the
full null sample, from~which percentiles and confidence intervals are obtained.
These intervals are percentile intervals of the null sample. They describe the
null distribution, not a confidence interval for the estimate. For~transfer
entropy, the~default effective value subtracts the estimate obtained from a
single resample. The~library also provides a multi-surrogate variant that
averages over several surrogates and reports their standard error. That standard
error describes the variability of the surrogate mean under source resampling,
not a joint-distribution confidence interval. The~tests are available wherever
permuting a source is meaningful, so calling one on a single-variable entropy
estimator is a compile error rather than a runtime failure. A~$p$-value measures significance and not
effect size, so the significance level and any correction for multiple
comparisons are left to the analyst. Seeded draws are reproducible and
independent of execution order, and~the optional parallel path returns identical
results. The~formal test definitions and the surrogate schemes are collected in
Appendix~\ref{app:tests}.

\subsection{Architecture and~API}

The library is organised around a small set of traits that separate the value a
measure yields (global, local/pointwise, or~conditional) from the estimator
that produces it. Each approach (discrete, kernel, ordinal, exponential family)
lives in its own module and implements the same interfaces, so measures are
composed rather than reimplemented, and~local values are derived consistently
across approaches. Constructors are provided as factory methods on the measure
types.

Inputs of arbitrary dimensionality are supported, with~the dimension handled at
compile time through const generics. Let us denote by $d$ the dimensionality of
an input variable, that is, the~number of coordinates per sample. Measures differ in
how many such variables they take (two for mutual information, three for
conditional mutual information and transfer entropy), and~each variable may have
its own $d$. Embedding raises the effective dimension an estimator works in:
delay embedding for transfer entropy, and~permutation patterns for the ordinal
estimators expand each sample into a vector whose dimension is a multiple of $d$,
so the joint space grows even when $d$ does not. The~estimators are optimised for
these higher-dimensional, sparse regimes in the computational sense, in~which the
probability density thins out and larger samples are needed for a stable estimate.
These runtime optimisations do not alter the estimator or the quantity it
computes. The~same mathematical quantity is evaluated, so the optimisations do
not trade accuracy or finite-sample bias for speed. This optimisation does not
improve the statistical properties of the estimator: the validation study in
Appendix~\ref{app:validation} shows the natural increase of the finite-sample
bias of the continuous estimators with dimension, so larger samples are required
even when the runtime remains~favourable.

The nearest-neighbour estimators inherit the dimension dependence of k-d trees.
In low dimension a balanced tree answers a single neighbour query in roughly
$O(\log N)$, so $N$ queries cost $O(N\log N)$. As~$d$ grows, the~bounding boxes
overlap, the~per-query cost approaches a linear scan, and~the number of points
needed for a stable estimate rises. The~library therefore switches strategy with
dimension and size: below the measured crossover it uses the exact k-d tree, and~above it the exponential-family estimators switch to a dense GPU calculation
(Section~\ref{sec:features}). The~parallel path splits the query work on the CPU.
The cross-package benchmark uses one-dimensional inputs, the~simplest common
setting supported by every package. Higher-dimensional and embedded inputs are
handled by the same estimator code, and~the test suite and
Appendix~\ref{app:scaling} exercise them. In~practice, increasing $d$ raises the
sample size required for a stable estimate and shifts the CPU/GPU crossover to
smaller $N$.

\subsection{GPU and Multi-Threaded~Paths}
\label{sec:features}

Two opt-in features accelerate computationally intensive estimators, both off
by~default.
\begin{itemize}
\item \texttt{Gpu} (wgpu, portable across Vulkan/Metal/DX12): kernel density
      evaluation for the kernel estimators, the~dense histogram for the
      plug-in discrete entropy, and~a dense nearest-neighbour path for the
      exponential-family entropy estimators (Kozachenko--Leonenko, R\'enyi and
      Tsallis). Dispatch is gated per estimator by sample size and dimension.
      The default gates are about 1200 points for the Gaussian kernel,
      4000 for the box kernel, and~4000 points with $D\ge8$ for the dense
      path. On~discrete GPUs the sample-size thresholds are lower, about
      1000 points for both kernels and 2000 points with $D\ge4$ for the
      dense path. Integrated GPUs, including Apple Silicon, dispatch at the
      thresholds above. In~both cases the GPU path is used once the threshold is
      met, and~only the crossover point differs. All gates are machine-relative
      and can be overridden with environment variables.
\item \texttt{Parallel} (rayon): parallelises the CPU kernel density loops and
      the neighbour-counting phases of the KSG family.
\end{itemize}

The cross-package comparison uses neither feature, so it is conservative and
hardware-independent. The~accelerated runs are reported separately in
Section~\ref{sec:perf}.

\subsection{Usage}

Listings~\ref{lst:entropy}--\ref{lst:te} show how to compute the discrete
entropy, the~KSG mutual information and the transfer entropy through the public
application programming interface (API). Local
(point-wise) values are illustrated on the mutual information but are provided
wherever the estimator supports them.

\begin{lstlisting}[caption={Discrete entropy.},label={lst:entropy},frame=single, xleftmargin=0.45em, xrightmargin=0.3em]
use infomeasure::estimators::entropy::Entropy;
use infomeasure::estimators::traits::{GlobalValue, LocalValues};
use ndarray::array;

let data = array![1, 0, 1, 1, 0, 1, 0, 0];
let h = Entropy::new_discrete(data).global_value(); // (*@$H = \lstEntropy\ldots$ nats@*)
\end{lstlisting}

\begin{lstlisting}[caption={KSG mutual information, global and local.},label={lst:mi},frame=single, xleftmargin=0.45em, xrightmargin=0.3em]
use infomeasure::estimators::mutual_information::MutualInformation;

let x = array![0.0, 1.0, 2.0, 3.0, 4.0, 5.0, 6.0, 7.0];
let y = array![0.0, 1.0, 2.0, 3.0, 4.0, 5.0, 6.0, 7.0];
let mi = MutualInformation::new_ksg(&[x, y], 4, 1e-10);
let mi_global = mi.global_value(); // (*@$I = \lstMi\ldots$ nats@*)
let mi_local = mi.local_values();  // one value per sample
\end{lstlisting}

\begin{lstlisting}[caption={Transfer entropy.},label={lst:te},frame=single, xleftmargin=0.45em, xrightmargin=0.3em]
use infomeasure::estimators::transfer_entropy::TransferEntropy;

let source = array![0, 1, 0, 1, 0, 1, 0, 1];
let dest = array![0, 0, 1, 0, 1, 0, 1, 0];
let te = TransferEntropy::new_discrete_mle(&source, &dest, 1, 1, 1);
let te_global = te.global_value(); // (*@$T = \lstTe\ldots$ nats@*)
\end{lstlisting}

\subsection{Documentation, Community, and~Availability}
\label{sec:meta}

This subsection collects the practical links and availability~information.
\begin{itemize}
\item Project landing page, language-independent, with~a guide to
      choosing between the Python and Rust packages and the interactive
     benchmarks:
 {\url{https://infomeasure.org/} (accessed on 23 November 2026)} 
\item Source code, issues, and~releases:
      {\url{https://codeberg.org/cbueth/infomeasure-rs} (accessed on 23 November 2026)}
\item API reference and guides, including a quick-start guide and
      guidance on choosing an estimator:
      {\url{https://docs.rs/infomeasure/latest/infomeasure/} (accessed on 23 November 2026)}
\item Crate, distributed on crates.io:
      {\url{https://crates.io/crates/infomeasure} (accessed on 23 November 2026)}
\item Contributing guidelines:
      {\url{https://codeberg.org/cbueth/infomeasure-rs/src/branch/main/CONTRIBUTING.md} (accessed on 23 November 2026)}
\item Code of conduct:
      {\url{https://codeberg.org/cbueth/infomeasure-rs/src/branch/main/CODE_OF_CONDUCT.md} (accessed on 23 November 2026)}
\end{itemize}

\section{Validation}
\label{sec:validation}

Correctness is established by an automated test suite run in continuous
integration, not by the illustrative examples of Section~\ref{sec:examples}.
The tests and their continuous-integration configuration are part of the public
repository (\url{https://codeberg.org/cbueth/infomeasure-rs} (accessed on 23 November 2026)).

The core of the strategy is parity testing: the same estimator, run on
the same data with the same parameters, must reproduce the Python reference
implementation to numerical tolerance. The~Python package was itself validated
against analytical solutions and across parameter grids
~\cite{buthInfomeasureComprehensivePython2025a}, so parity transfers that
validation to the Rust implementation and catches any divergence introduced by
optimisation. On~top of parity, the~suite contains independent validating tests:
analytically solvable sanity cases (for example, a~Gaussian variable with known
differential entropy, or~a bivariate Gaussian with known mutual information),
unit and utility tests, and~edge cases such as degenerate inputs and tied
distances.
At the time of writing, the~suite comprises more than 1600 tests, including
more than 1000 parity tests, analytical sanity tests, and~unit and utility
tests. The~CPU test suite spans inputs of up to 8 dimensions, and~the GPU
test path up to 32, including multivariate configurations. The~revision adds a
broader analytical validation study against Gaussian and linear-Gaussian-VAR
ground truth in Appendix~\ref{app:validation}.

\section{Performance~Evaluation}
\label{sec:perf}
\unskip

\subsection{Methodology}

Every package is run on exactly the same data, with~the same parameters, on~a
single CPU core, and~only the computation of the measure itself is timed.
Reading the data and starting the program are outside the timer. Warm-up runs
are discarded, each setting is repeated several times, and~we report the average
runtime and its spread. No package is given an advantage: \textit{infomeasure-rs}
uses neither its GPU nor its multi-threaded mode here, and~every package is
measured with the same~procedure.

Results are compared through a per-configuration speed-up factor. For~each
configuration $i$, comprising a fixed measure, estimator, and~input, let
$t^{\text{rs}}_i$ and $t^{\text{other}}_i$ be the measured runtimes of
\textit{infomeasure-rs} and of the competing package on identical inputs. We
call the ratio of the two the speed-up $S_i$,
\begin{equation}
S_i = \frac{t^{\text{other}}_i}{t^{\text{rs}}_i},
\qquad\text{so that}\quad S_i > 1 \iff \text{\textit{infomeasure-rs} is faster.}
\end{equation}
Runtimes
 span several orders of magnitude and the differences we care about are
relative, so we do not average the speed-ups $S_i$ arithmetically. Across the
$n$ configurations we report their geometric mean,
\begin{equation}
\overline{S}_{\text{geo}}
= \left(\prod_{i=1}^{n} S_i\right)^{1/n}
= \exp\!\left(\frac{1}{n}\sum_{i=1}^{n}\ln S_i\right),
\end{equation}
the typical factor by which a competing package is slower and, on~the
logarithmic axes used below, the~midpoint of the $S_i$. Confidence intervals for these summaries are obtained
on the same log scale,
$\exp\!\left(\overline{\ln S} \pm 1.96\,\mathrm{sd}(\ln S)/\sqrt{n}\right)$,
that is, by~propagating the spread of the per-configuration speed-ups. Wherever
a single number summarises a comparison, we give the geometric-mean speed-up
$\overline{S}_{\text{geo}}$ and we also give the fraction of configurations in
which \textit{infomeasure-rs} is~faster. ({Averaging ratios
 is multiplicative: a configuration in which
\textit{infomeasure-rs} is twice as fast as the compared package ($S_i=2$) and
one in which it is twice as slow ($S_i=\tfrac12$) carry equal and opposite
weight and cancel to $\overline{S}_{\text{geo}}=1$, i.e.\ no overall advantage,
as $\operatorname{GM}(2,\tfrac12)=1$ (with $\operatorname{GM}$ the geometric
mean), whereas the arithmetic mean would report a spurious $1.25$. The~ratio scale is scale-free, since multiplying both runtimes
by the same factor leaves every $S_i$ unchanged, and~the mean is invertible,
$\operatorname{GM}(1/S_i)=1/\operatorname{GM}(S_i)$: a result may be quoted as a
speed-up ($S>1$ faster) or as the reciprocal runtime ratio ($<$1 faster) without
changing the quantity.})

The intervals are descriptive summaries over configurations, not iid confidence
intervals for a population parameter. For~significance we use a per-configuration
Welch test on the reported means and standard deviations, one-sided in the
direction of the observed difference, and~apply Holm correction across the
matched representative pairs. For~package-level summaries we also report a paired
sign-flip test on the configuration-level log speed-ups and a bootstrap interval
over configurations. The~per-configuration tests are the inferential unit
because each package shares a different set of~configurations.

Concretely, the~runs use byte-identical datasets of $N$ samples generated once
with fixed seeds and identical parameters. The~benchmark inputs are
one-dimensional, that is, $d=1$. Each configuration follows the same adaptive
schedule:
up to three warm-up calls bound to a 0.4\,s budget, then three to ten timed
iterations bound to 1.5\,s per draw. The~mean and standard deviation are pooled
across the random input draws, which are identical for every package. The~representative configurations use four draws and the remaining parameter sweeps
a single draw. Thread counts are pinned everywhere: JIDT \texttt{NUM\_THREADS = 1};
the numerical libraries \texttt{OMP}, \texttt{OPENBLAS}, \texttt{MKL},
\texttt{NUMEXPR}, \texttt{RAYON} and \texttt{VECLIB} set to one; and
Python~3.12 with \texttt{workers = 1}. The compared versions were infomeasure
(Python) 0.6.3, infomeasure-rs 0.5.0, JIDT 1.6.1, PyInform 0.2.0, pyitlib 0.3.1,
dit 2.2, Syntropy 0.0.2, DiscreteEntropy.jl 0.2.1, pyEntrp 2.1.0,
RTransferEntropy 0.2.21, logp 0.2.3, and entropium 0.2.0. Julia 1.12.7 and
OpenJDK 21 were used for the Julia and Java packages. One shared harness,
part of the public repository, drives every package and reports the mean,
standard deviation, median, confidence interval, and~sample count, so no library
is measured with a private procedure that could flatter it. Third-party semantic
differences (for example, a~quantile-binned transfer-entropy estimator, or~a
package that builds empirical distributions inside the timed call) are recorded
as~caveats.

We report these equalised, single-thread CPU numbers throughout. The~crate also
ships two opt-in accelerators that make the same estimators faster still and are
not part of this comparison: a portable GPU path (wgpu) for the large-sample
paths where a GPU is available, and~a multi-threaded CPU path (rayon) that
covers the same loops where no GPU applies. They are complementary by design and
are never combined in one call. All benchmark numbers, including the full size
sweep, are also available in an interactive view at
\url{https://infomeasure.org/benchmarks/} (accessed on 23 November 2026).

\subsection{Comparison with Other~Packages}

\begin{figure}[H]
\includegraphics[width=\textwidth]{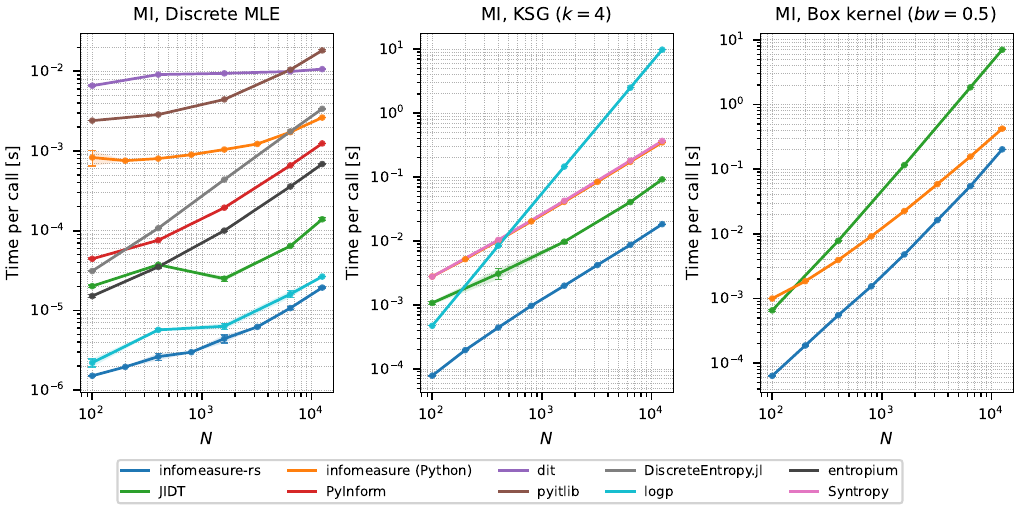}
\caption{Mutual
 information runtime versus sample size, on~a single thread, for~every package in the benchmark that provides the estimator: from left to right
the discrete plug-in estimator, the~KSG estimator with $k=4$, and~the box kernel
with bandwidth $0.5$. Packages are colours and the shaded bands are 95\%
confidence intervals of the mean. For~the middle plot, \textit{infomeasure} (Python) and Syntropy are overlapping. \label{fig:mi}}
\end{figure}

\begin{figure}[H]
\includegraphics[width=0.95\textwidth]{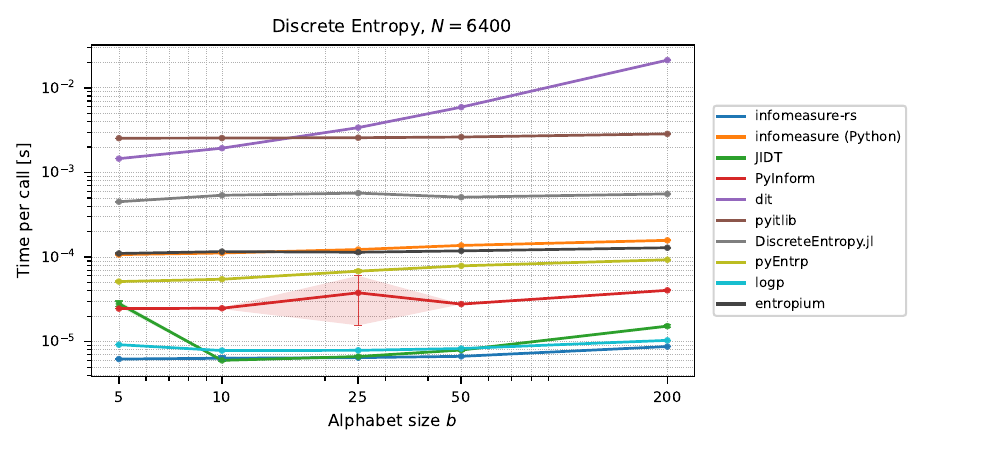}
\caption{Discrete
entropy runtime versus alphabet size at $N=6400$, for~every
package in the benchmark that provides it, on~a single thread. Each package is
a colour. The~shaded bands are the 95\% confidence intervals of the mean. This
is the one view where a competitor is marginally ahead, with~JIDT for alphabet sizes
around ten to twenty-five. \label{fig:alphabet}}
\end{figure}

Table~\ref{tab:cross} reports the geometric-mean speed-up of
\textit{infomeasure-rs} over each third-party package for the measures they
share, using the representative parameter slices (\texttt{k}$=4$,
bandwidth $=0.5$, one method per family). A~missing cell means that a package
does not expose that measure for that estimator directly, not that the measure
cannot be derived from entropy or cross-entropy sums. The~library exposes all
measures and implements the corresponding estimators, rather than only the
simple sums. Each row is therefore a fair comparison for that pair, but~the rows
are not a like-for-like ranking of the third-party packages, because~the shared
configuration sets differ. The~comparison set spans the Python,
C, Java, Julia and R packages identified in the companion paper's survey,
rather than a convenient subset, and~the library is faster than every one of
them on the shared measures. The~two Rust crates from
Table~\ref{tab:rustsurvey} that implement directly comparable measures,
\texttt{logp} and \texttt{entropium}, are included here. The~other Rust crates
in that table are absent because they offer no estimator that can be timed
fairly on these measures, being single-purpose, stale, or~outside the scope of
the comparison (Section~\ref{sec:rust-survey}). \textit{infomeasure-rs} is the
fastest package in \sigFaster{} of the \sigTotal{} matched representative
configurations (\sigFasterPct\%) and at parity in the remaining \sigPar{}
(\sigParPct\%). No competitor is significantly faster in any configuration.
The few close cases are informative: for the most common measure, discrete
Shannon entropy, JIDT and \texttt{logp} approach parity at particular alphabet
and sample sizes, and~the Welch tests do not reject the null of equal runtime in
those cases.
For each configuration, the~Welch test with Holm correction finds
\textit{infomeasure-rs} significantly faster in \sigFaster{} cases, not
distinguishable in \sigPar{}, and~significantly slower in \sigSlower{}.
Figure~\ref{fig:mi} shows the scaling with sample size for
mutual information across every package that provides each estimator, and~Figure~\ref{fig:alphabet} gives the entropy view as a function of alphabet size
at ${N} =6400$
. For~the conditional and transfer measures (CMI, TE and CTE) and
for the KSG, kernel, ordinal and generalised-entropy estimators the margins are
much larger. Some of these optimisations were motivated by noticing that a
competing package was faster for a particular estimator. The~0.4.x and 0.5.0
releases fold in several such improvements, although~the library was already
competitive before~them.

\unskip

\begin{landscape} 
\begin{table}[H]
\centering
\caption{Geometric-mean
 speed-up of \textit{infomeasure-rs} over each package,
$t_{\text{package}}/t_{\text{rs}}$ (${>}1$ means the Rust library is faster),
over the representative entries they share. Columns are grouped by measure
(top row) and estimator family (bottom row: discrete, nearest-neighbour, box
kernel, Gaussian kernel). A~dash means the package does not expose that measure
for that estimator directly. Measures derived from entropy or cross-entropy
sums are not counted, because~the comparison is restricted to what each package
provides as a direct estimator. Every aggregate cell is at or above parity.
Significance is assessed with per-configuration Welch tests and Holm correction.
Rows are not directly comparable because each package shares a different
configuration set.
\label{tab:cross}}
\footnotesize
\setlength{\tabcolsep}{2pt}
\resizebox{\linewidth}{!}{%
\begin{tabular}{ccccccccccccccccccccc}

\toprule
 & \multicolumn{4}{c}{\textbf{Entropy}} & \multicolumn{4}{c}{\textbf{MI}} & \multicolumn{4}{c}{\textbf{CMI}} & \multicolumn{4}{c}{\textbf{TE}} & \multicolumn{4}{c}{\textbf{CTE}} \\
\cmidrule{2-21}
\textbf{Package} & {\textbf{Disc.}} & {\textbf{KL}} & {\textbf{Box}} & {\textbf{Gauss}} & {\textbf{Disc.}} & {\textbf{KSG}} & {\textbf{Box}} & {\textbf{Gauss}} & {\textbf{Disc.}} & {\textbf{KSG}} & {\textbf{Box}} & {\textbf{Gauss}} & {\textbf{Disc.}} & {\textbf{KSG}} & {\textbf{Box}} & {\textbf{Gauss}} & {\textbf{Disc.}} & {\textbf{KSG}} & {\textbf{Box}} & {\textbf{Gauss}} \\
\midrule
infomeasure (Python) & 50.9 & 3.34 & 3.58 & 1.68 & 244 & 22.9 & 5.00 & 1.68 & 120 & 32.4 & 7.79 & 1.67 & 375 & 7.55 & 7.86 & 1.69 & 282 & 7.67 & 11.9 & 1.64 \\
JIDT & 7.44 & 34.5 & 24.9 & {--} & 8.58 & 6.41 & 21.0 & {--} & 3.66 & 5.10 & {--} & {--} & 7.21 & 4.44 & 11.0 & {--} & 4.71 & 4.90 & {--} & {--} \\
PyInform & 9.21 & {--} & {--} & {--} & 43.0 & {--} & {--} & {--} & 7.78 & {--} & {--} & {--} & 11.2 & {--} & {--} & {--} & 10.9 & {--} & {--} & {--} \\
Syntropy & {--} & 2.80 & {--} & {--} & {--} & 23.6 & {--} & {--} & {--} & 18.2 & {--} & {--} & {--} & {--} & {--} & {--} & {--} & {--} & {--} & {--} \\
logp & 1.13 & {--} & {--} & {--} & 1.56 & 66.3 & {--} & {--} & {--} & {--} & {--} & {--} & {--} & {--} & {--} & {--} & {--} & {--} & {--} & {--} \\
pyitlib & 652 & {--} & {--} & {--} & 1098 & {--} & {--} & {--} & 324 & {--} & {--} & {--} & {--} & {--} & {--} & {--} & {--} & {--} & {--} & {--} \\
DiscreteEntropy.jl & 55.4 & {--} & {--} & {--} & 75.1 & {--} & {--} & {--} & {--} & {--} & {--} & {--} & {--} & {--} & {--} & {--} & {--} & {--} & {--} & {--} \\
dit & 727 & {--} & {--} & {--} & 1749 & {--} & {--} & {--} & {--} & {--} & {--} & {--} & {--} & {--} & {--} & {--} & {--} & {--} & {--} & {--} \\
entropium & 13.0 & {--} & {--} & {--} & 20.5 & {--} & {--} & {--} & {--} & {--} & {--} & {--} & {--} & {--} & {--} & {--} & {--} & {--} & {--} & {--} \\
RTransferEntropy & {--} & {--} & {--} & {--} & {--} & {--} & {--} & {--} & {--} & {--} & {--} & {--} & 102 & {--} & {--} & {--} & {--} & {--} & {--} & {--} \\
pyEntrp & 23.0 & {--} & {--} & {--} & {--} & {--} & {--} & {--} & {--} & {--} & {--} & {--} & {--} & {--} & {--} & {--} & {--} & {--} & {--} & {--} \\
\bottomrule
\end{tabular}}

\end{table}
\end{landscape}

\unskip

\subsection{Comparison with the Python~Reference}

Across \matchedPairs\ matched configurations, \textit{infomeasure-rs} is a
geometric-mean \num{\geomeanSpeedup \pm \geomeanSpeedupPm} times faster than the
Python reference on a single thread; that is, it needs
\num{\geomeanVsPython \pm \geomeanVsPythonPm} of the reference runtime, and~it is
faster in \fasterPct\% of them. The~uncertainties are 95\% confidence intervals. Table~\ref{tab:rustpy} breaks the
speed-up down by estimator family and measure. The~discrete and ordinal
families show the largest margins, while the kernel family, whose reference
implementation evaluates the density in compiled Fortran, is the~closest.

\subsection{GPU and Multi-Threaded~Acceleration}

\begin{figure}[H]
\includegraphics[width=0.72\textwidth]{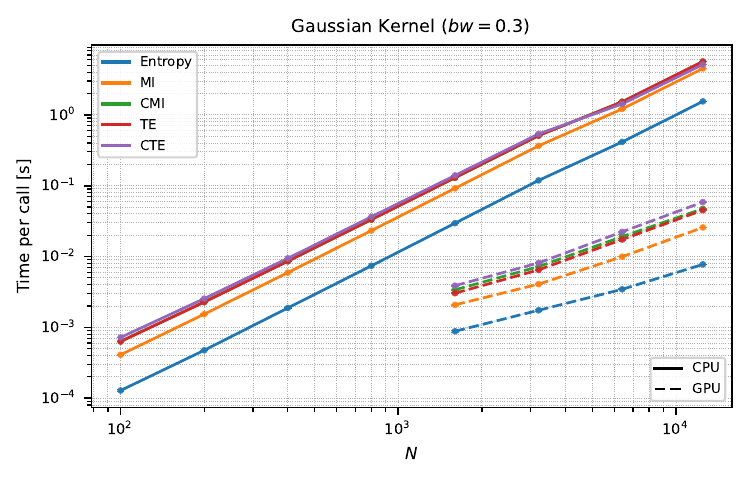}
\caption{Gaussian-kernel
 runtime ($bw=0.3$) for \textit{infomeasure-rs} on the
CPU (solid) and GPU (dashed), one colour per measure, with~95\% confidence
bands. The~CMI and TE estimators have almost the same runtime on this hardware,
so their curves overlap. Enabling the GPU makes it a further factor of
\gpuSpeedupMin{} to \gpuSpeedupMax{} faster than the single-thread CPU path. The~cross-package comparison in this section uses neither the GPU nor
multi-threading, so these gains are optional and additional to an already
favourable baseline.
\label{fig:gpu}}
\end{figure}

When the \texttt{gpu} feature is enabled, the~kernel estimators are accelerated
further once the sample size exceeds a machine-dependent gate. Figure~\ref{fig:gpu}
shows the Gaussian-kernel case. The~\texttt{parallel} feature accelerates the
kernel density loops and the KSG neighbour counting on the CPU. These are
additional to the single-thread CPU result and are not used in
Tables~\ref{tab:cross} and \ref{tab:rustpy}.

\begin{table}[H]
\caption{Geometric-mean
 speed-up $t_{\text{Python}}/t_{\text{Rust}}$
(${>}1$ means the Rust library is faster) by estimator and measure, on~a single
thread. \label{tab:rustpy}}
\centering
{\small

\begin{tabularx}{\textwidth}{lCCCCC}

\toprule
\textbf{Estimator} & \multicolumn{1}{c}{\textbf{Entropy}} & \multicolumn{1}{c}{\textbf{mi}} & \multicolumn{1}{c}{\textbf{cmi}} & \multicolumn{1}{c}{\textbf{te}} & \multicolumn{1}{c}{\textbf{cte}} \\
\midrule
Discrete & 26.0 & 46.9 & 32.7 & 49.4 & 34.8 \\
KSG & 3.25 & 21.9 & 29.0 & 7.64 & 8.04 \\
Ordinal & 33.5 & 84.0 & 75.6 & 645 & 610 \\
Kernel (box) & 3.54 & 4.91 & 7.61 & 7.66 & 11.3 \\
Kernel (Gaussian) & 1.65 & 1.66 & 1.67 & 1.68 & 1.67 \\
R\'enyi & 2.58 & 2.56 & 2.41 & 2.40 & 2.33 \\
Tsallis & 2.55 & 2.55 & 2.40 & 2.39 & 2.32 \\
\bottomrule
\end{tabularx}

}
\end{table}

Appendix~\ref{app:scaling} reports a separate dimensional scaling run. At~$N=5000$ and $d=32$ the GPU path reduces KL entropy from 224.0\,ms to 17.5\,ms,
and the parallel path reduces KSG mutual information from 667.4\,ms to
107.6\,ms. The~Python reference takes 408.8\,ms and 1966\,ms on the same inputs.
These numbers come from a separate scaling run on an Apple M4 Pro and are not
part of the cross-package~benchmark.

\begin{figure}[H]
\includegraphics[width=\textwidth]{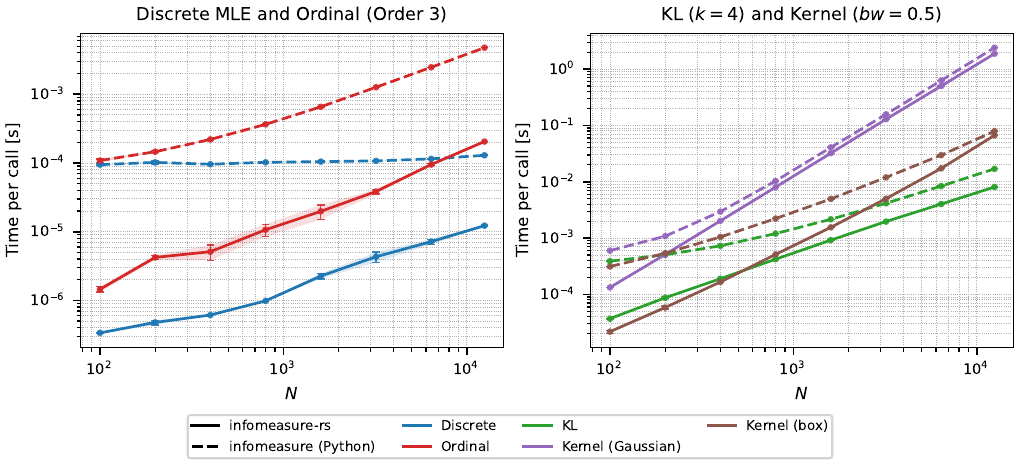}
\caption{Entropy
 runtime versus sample size for \textit{infomeasure-rs}
(solid) and the Python reference (dashed), on~a single thread. \textbf{Left} panel:
discrete plug-in and ordinal (order 3). \textbf{Right} panel: Kozachenko--Leonenko (KL)
with $k=4$ and the Gaussian and box kernels with bandwidth $0.5$. The~estimator is encoded by
colour and the package by line style. Shaded bands are 95\% confidence
intervals of the mean. They are so narrow that they are barely visible at this
scale. \label{fig:scaling}}
\end{figure}
\unskip

\section{Illustrative~Examples}
\label{sec:examples}

This section is illustrative rather than a validation, the~latter being the
parity suite of Section~\ref{sec:validation}. Its purpose is to show the library
in use on small, fully reproducible systems whose behaviour is familiar from
the literature and to exercise estimators beyond the plug-in discrete case.
Every value was computed with \textit{infomeasure-rs}. Each caption reports the
wall time of the measure calls only, excluding the generation of the system. The~full parameter sets are listed in Appendix~\ref{app:example-params}.

\subsection{Logistic Map (Entropy)}

\begin{figure}[H]
\includegraphics[width=0.56\textwidth]{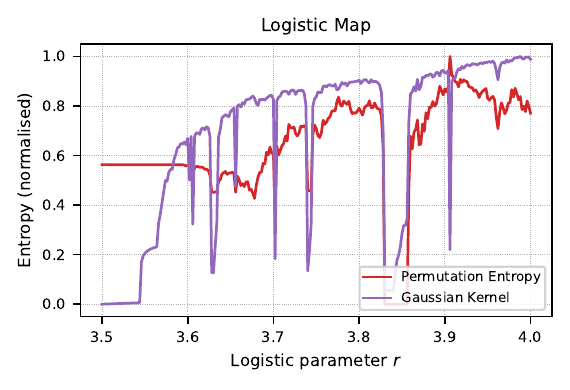}
\caption{Logistic map: order-3 permutation entropy and Gaussian-kernel entropy
versus the parameter $r$. Both curves are
min--max normalised to $[0,1]$ over the parameter range, since the two
estimators have different offsets. This makes their shapes directly
comparable. The~measure calls took \exLogisticSeconds~s for
\exLogisticCalls\ evaluations on an Apple M4 Pro. The~trajectory generation is
excluded. \label{fig:ex-logistic}}
\end{figure}

The logistic map $x_{t+1}=r\,x_t(1-x_t)$ \cite{may1976logistic} is the textbook
route to chaos, and~its entropy as a function of $r$ is a standard test case
for both permutation and continuous estimators~\cite{bandt2002permutation,cao2004permutation}. Figure~\ref{fig:ex-logistic}
computes the order-3 permutation entropy and the Gaussian-kernel entropy along
the bifurcation diagram. The~drops in the permutation entropy mark the periodic
windows, visible without any~tuning.

\unskip

\subsection{Kuramoto Oscillators (MI and TE)}

\begin{figure}[H]
\includegraphics[width=0.56\textwidth]{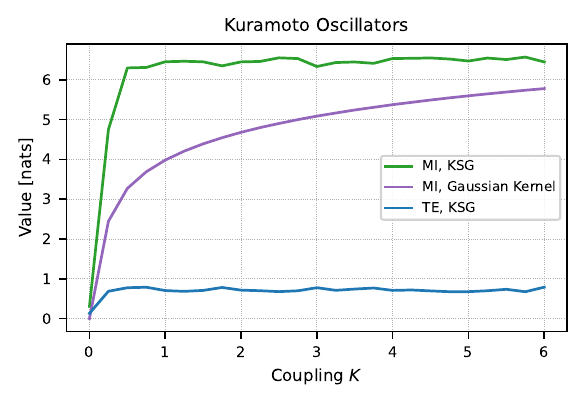}
\caption{Kuramoto oscillators: mutual information estimated with KSG and with a
Gaussian kernel, and~KSG transfer entropy, between~two oscillators as a
function of the coupling $K$. Each point is a single deterministic trajectory,
so no confidence interval is shown. The~measure calls took
\exKuramotoSeconds~s for \exKuramotoCalls\ evaluations on an Apple M4 Pro. The~integration is excluded. \label{fig:ex-kuramoto}}
\end{figure}

A population of $N=24$ phase oscillators with heterogeneous natural frequencies
and mean-field coupling $K$ \cite{strogatz2000kuramoto} undergoes a
synchronisation transition. Figure~\ref{fig:ex-kuramoto} tracks the mutual
information~\cite{miKSG2004} and the transfer entropy~\cite{schreiber2000measuring} between two oscillators as $K$ grows: the
information shared rises through the transition, and~the estimators behave
consistently in the ordered phase. The~absolute mutual-information values differ
between the two estimators because the Gaussian kernel depends on its bandwidth
and the KSG estimate carries a finite-sample bias. The~qualitative rise through
the transition is the same, which is the point of this~illustration.

\subsection{Coupled R\"ossler Systems (TE, R\'enyi and Tsallis)}

\begin{figure}[H]
\includegraphics[width=\textwidth]{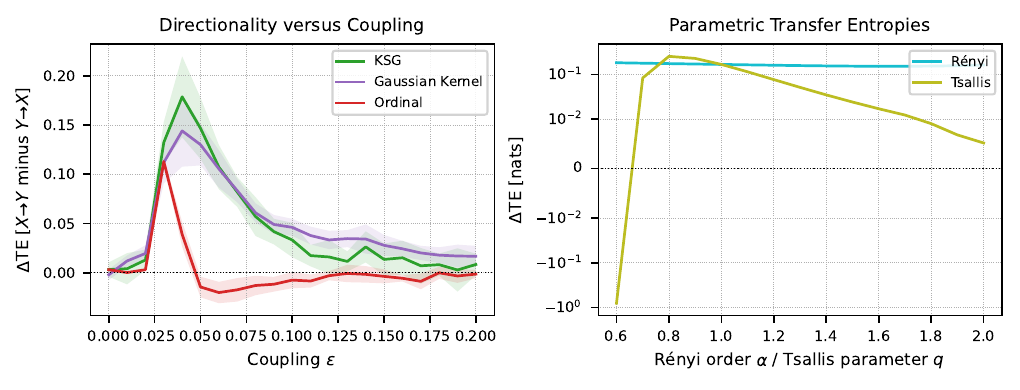}
\caption{Coupled R\"ossler systems. \textbf{Left}: forward-minus-backward transfer
entropy versus the coupling $\varepsilon$ for three estimators, averaged over ten
trajectories. Shaded bands are one standard deviation across trajectories.
\textbf{Right}: the same signed difference for the R\'enyi and Tsallis families on a
single trajectory versus their parameter (symmetric-log scale). A~positive value
indicates the correct causal direction. The~measure calls took
\exRosslerEpsSeconds~s
(\exRosslerEpsCalls\ evaluations, coupling sweep), \exRosslerRenyiSeconds~s
(\exRosslerRenyiCalls, R\'enyi sweep) and \exRosslerTsallisSeconds~s
(\exRosslerTsallisCalls, Tsallis sweep) on an Apple M4 Pro. The~integration is
excluded. \label{fig:ex-rossler}}
\end{figure}

Two unidirectionally coupled R\"ossler systems~\cite{rossler1976equation} are a
standard benchmark for causal inference, and~a case where the coupling is
carried mainly by large excursions rather than the bulk of the distribution~\cite{jizba2022causal,paluvs2024causes}. Figure~\ref{fig:ex-rossler} reports the
forward and backward transfer entropy obtained with the KSG, Gaussian-kernel
and ordinal estimators as a function of the coupling, together with the
parametric R\'enyi~\cite{renyi1961measures} and Tsallis
~\cite{tsallis1988possible} transfer entropies. We report the signed difference
$\Delta\mathrm{TE}$ between the forward and backward estimates, which is
positive when the coupling is correctly directed and is easier to read than
overlaying two near-identical curves. The~KSG and Gaussian-kernel estimates
indicate the correct direction across all couplings. The~ordinal estimate is
systematically lower and becomes negative for most couplings above
$\varepsilon\approx0.04$, reflecting its finite-sample bias and its lower
sensitivity to the large excursions that carry the coupling in this system. The~advanced estimators expose the dependence on $\alpha$ and $q$ that the Shannon
estimators~cannot.

\unskip

\unskip

\section{Discussion and~Conclusions}

This note has introduced \textit{infomeasure-rs}, validated it against the
Python reference and analytic cases, and~benchmarked it against the established
implementations. The~library brings the measure and estimator coverage of the
\textit{infomeasure} Python package to Rust, including the composite
Kullback--Leibler and Jensen--Shannon divergences, and~keeps the numerical
results identical through continuous parity testing. This lowers the four
barriers: estimator choice (one dimension-generic
interface), fragmentation (one package across measures and estimators),
reproducibility (one implementation shared across labs and fields instead of
throwaway code), and~computational cost (speed). On~a fair single-thread,
identical-input benchmark it is the fastest of the packages we compared,
matching the closest ones only for plug-in discrete entropy and pulling
clearly ahead for mutual information, the~conditional and transfer measures,
and the kernel, ordinal and generalised-entropy estimators. Optional GPU and
multi-threaded paths accelerate selected continuous estimators further, and~the
library compiles to self-contained native code. To~our knowledge it is the most
complete information-theory package for Rust. All benchmark numbers, including
the full size sweep, are also available in an interactive view at
\url{https://infomeasure.org/benchmarks/} (accessed on 23 November 2026).

The limitations follow directly from the design. The~comparison is runtime-only
and does not assert accuracy, which the validation suite covers separately. The~validation study also shows that the finite-sample bias of the continuous
estimators grows with dimension, so the runtime advantage does not remove the
need for larger samples. It
spans the comparable packages we could identify across languages rather than an
exhaustive census, and~the cross-package benchmark uses one-dimensional inputs, with~higher dimensions exercised by the test suite and separate benchmarks. GPU
acceleration is device-dependent and covers the kernel estimators, the~dense
plug-in discrete histogram, and~the dense nearest-neighbour path for the
exponential-family entropy estimators. The~counting-based KSG estimators stay on
the CPU because the optimised CPU path is faster. The~Rust ecosystem is
young and the API is type-safe rather than dynamically selectable, a~deliberate
trade-off.
\vspace{6pt}
\authorcontributions{Conceptualization, methodology, software, validation,
formal analysis, investigation, data curation, writing---original draft
preparation, writing---review and editing, and~visualization: C.M.B. The~author has read and agreed to the published version of the~manuscript.}

\funding{This research received no external funding. The~author carried out
this work in his own time and in a personal~capacity.}

\institutionalreview{Not applicable.}
\informedconsent{Not applicable.}

\dataavailability{The benchmark measurements and the harness that collects them
are available in the project repository and on the project's benchmark page
(\url{https://infomeasure.org/benchmarks/} (accessed on 23 November 2026)). The~snapshot used here is the
\texttt{pages} commit
\href{https://codeberg.org/cbueth/infomeasure-rs/src/commit/7c73725}{\texttt{7c73725}},
produced with version~0.5.0 and collected by
\path{scripts/bench_packages/run_all.sh} through
\path{.woodpecker/bench-packages.yml}, which also checks for upstream package
updates. The~timed code paths are unchanged by the later statistical additions.
The software is archived on Zenodo, with version DOI
\url{https://doi.org/10.5281/zenodo.24474818} for version~0.7.0 and concept DOI
\url{https://doi.org/10.5281/zenodo.23477339} for the latest version. The~project landing page is \url{https://infomeasure.org/} (accessed on 23 November 2026). The~source code is
available at \url{https://codeberg.org/cbueth/infomeasure-rs} (accessed on 23 November 2026), documented at
\url{https://docs.rs/infomeasure/latest/infomeasure/} (accessed on 23 November 2026), and~distributed as the
crate \url{https://crates.io/crates/infomeasure} (accessed on 23 November 2026).}

\acknowledgments{The author thanks the \textit{infomeasure} collaborators for
discussions and support and Massimiliano Zanin for proofreading the
manuscript. During
 the preparation of this manuscript the
author used OpenCode, model deepseek-v4.1-flash, to~assist with
editing text and plotting. Benchmark data was collected independently.
No other large language model tool was used. The~author reviewed and edited all content and takes full
responsibility for the work.}

\conflictsofinterest{The author declares no conflicts of~interest.}

\abbreviations{Abbreviations}{
The following abbreviations are used in this manuscript:\\

\vspace{3pt}

\noindent
\begin{tabular}{@{}ll}
CMI & Conditional mutual information\\
CTE & Conditional transfer entropy\\
GPU & Graphics processing unit\\
H & Entropy\\

JSD & Jensen--Shannon divergence\\
KL & Kozachenko--Leonenko\\

KLD & Kullback--Leibler divergence\\
kNN & $k$-nearest neighbour\\
KSG & Kraskov--St\"ogbauer--Grassberger\\
MI & Mutual information\\
MLE & Maximum likelihood (plug-in)\\
NSB & Nemenman--Shafee--Bialek\\
TE & Transfer entropy\\
\end{tabular}
}

\appendix
\setcounter{table}{0}
\setcounter{figure}{0}
\setcounter{equation}{0}
\renewcommand{\thetable}{A\arabic{table}}
\renewcommand{\thefigure}{A\arabic{figure}}
\renewcommand{\theequation}{A\arabic{equation}}

\section{Measures, Estimators, and~Limits}
\label{app:definitions}

Throughout this appendix \(X\), \(Y\) and \(Z\) denote random variables, with~\(Z\) collecting the conditioning variables, \(p\) and \(q\) denote probability
mass or density functions, and~logarithms are natural. Hats denote~estimators.

For a discrete distribution $p(x)$ the Shannon entropy is
\begin{equation}
H(X)=-\sum_x p(x)\log p(x).
\end{equation}
The same expressions apply to continuous variables with sums replaced by
integrals. Mutual information, conditional mutual information, transfer entropy,
and conditional transfer entropy are defined through entropy sums,
\begin{align}
I(X;Y)&=H(X)+H(Y)-H(X,Y),\\
I(X;Y\mid Z)&=H(X,Z)+H(Y,Z)-H(X,Y,Z)-H(Z),\\
T_{X\to Y}&=I(X_{\mathrm{past}};Y_{\mathrm{future}}\mid Y_{\mathrm{past}}),\\
T_{X\to Y\mid Z}&=I(X_{\mathrm{past}};Y_{\mathrm{future}}\mid Y_{\mathrm{past}},Z_{\mathrm{past}}),
\end{align}
with the past and future blocks formed by the delay embedding described in
Section~\ref{sec:package}. The~Kullback and Leibler divergence and the Jensen and
Shannon divergence are
\begin{align}
D_{\mathrm{KL}}(P\parallel Q)&=H_Q(P)-H(P),\\
\mathrm{JSD}_{\pi}(P_1,\dots,P_n)&=H\!\left(\sum_i \pi_i P_i\right)-\sum_i \pi_i H(P_i),
\end{align}
where $P$, $Q$ and $P_1,\dots,P_n$ are probability distributions,
$\pi_1,\dots,\pi_n$ are weights with $\sum_i\pi_i=1$, and~$H_Q(P)=-\sum_x p(x)\log q(x)$ is the cross-entropy of \(P\) relative to
\(Q\).

The R\'enyi and Tsallis families are
\begin{align}
H_{\alpha}(X)&=\frac{1}{1-\alpha}\log\sum_x p(x)^{\alpha},\qquad \alpha>0,\ \alpha\ne1,\\
H_{q}(X)&=\frac{1}{q-1}\left(1-\sum_x p(x)^{q}\right),\qquad q>0,\ q\ne1.
\end{align}
Both reduce to the Shannon entropy in the limits $\alpha\to1$ and $q\to1$. The~generalised mutual information, conditional mutual information, transfer entropy,
and conditional transfer entropy use the same entropy-sum construction with
$H_{\alpha}$ or $H_{q}$ in place of $H$. Tsallis mutual information has also been
used to build mutual-information matrices and global measures for multivariate
time series~\cite{behera2025mutual}. The~Tsallis
entropy is non-additive. For~two independent systems $A$ and $B$, with~\(A+B\) their composite,
\begin{equation}
H_q(A+B)=H_q(A)+H_q(B)+(1-q)H_q(A)H_q(B),
\end{equation}
so the entropy of a composite system is not the sum of its parts unless $q=1$.

For continuous variables, the~nearest-neighbour R\'enyi and Tsallis estimators use
the $k$-th neighbour distance $\rho_{i,k}$,
\begin{align}
\hat H_{\alpha}(X)&=\frac{1}{1-\alpha}\log\left[\frac{1}{N}\sum_{i=1}^{N}
\left((N-1)C_{k,\alpha}V_m\rho_{i,k}^{m}\right)^{1-\alpha}\right],\\
C_{k,\alpha}&=\left[\frac{\Gamma(k)}{\Gamma(k+1-\alpha)}\right]^{1/(1-\alpha)},\\
\hat H_{q}(X)&=\frac{1-\hat I_q}{q-1},\qquad
\hat I_q=\frac{1}{N}\sum_{i=1}^{N}
\left((N-1)C_{k,q}V_m\rho_{i,k}^{m}\right)^{1-q},
\end{align}
where \(N\) is the sample size, \(i\) indexes the sample, \(V_m\) is the volume
of the \(m\)-dimensional unit ball, \(m\) is the embedding dimension, and~\(\Gamma\) is the gamma function. \(C_{k,q}\) is defined as \(C_{k,\alpha}\)
with \(\alpha\) replaced by \(q\). The~estimator requires \(1\le k<N\) and
$\alpha<k+1$ for R\'enyi or $q<k+1$ for Tsallis, with~the Shannon limit handled
separately. These restrictions follow from the gamma correction~factor.

For the generalised families the cross-entropy is defined so that it reduces to
$H_Q(P)$ in the Shannon limit. For~discrete distributions,
\begin{align}
H^{\alpha}_Q(P)&=\frac{1}{1-\alpha}\log\sum_x p(x)\,q(x)^{\alpha-1},\\
H^{q}_Q(P)&=\frac{1-\sum_x p(x)\,q(x)^{q-1}}{q-1}.
\end{align}
The generalised KLD is the generalised cross-entropy minus the generalised
entropy,
\begin{equation}
D^{\alpha}_{\mathrm{KL}}(P\parallel Q)=H^{\alpha}_Q(P)-H_{\alpha}(P),
\end{equation}
and likewise for $q$. As~$\alpha\to1$ or $q\to1$ the cross-entropy tends to
$H_Q(P)=-\sum_x p(x)\log q(x)$ and the divergence tends to the Shannon KLD. The~continuous estimators use the same construction, with~the neighbour distances
measured against the reference sample $Q$ instead of the sample $P$.

A generalised JSD would need the generalised entropy of the mixture
$\sum_i\pi_iP_i$. That entropy is not fixed by the generalised entropies of the
components, so the Jensen and Shannon identity does not carry over to the
generalised families. The~library provides JSD for the discrete, kernel and
ordinal estimators, which expose a normalised mass function for the mixture, with~a pooled kernel density in the continuous case. The~R\'enyi and Tsallis families
and the nearest-neighbour estimator do not provide one, so JSD is not available
for them. This is the reason for the empty cells in
Table~\ref{tab:coverage}.

Except for the nearest-neighbour R\'enyi and Tsallis forms above, the~expressions
in this appendix define the measures rather than the estimators. The~library
evaluates them with dedicated bias-reduced estimators, including the
Kozachenko--Leonenko, kernel, ordinal and bias-corrected discrete estimators.
These are documented in the crate guide. They target the same quantities and are
algebraically equivalent to the definitions above but have lower finite-sample
bias.

\section{Statistical Tests and Surrogate~Schemes}
\label{app:tests}

Let $\mathcal{T}$ be the measure and $t$ its observed value. A~permutation test
builds $B$ resampled values $\mathcal{T}_b=\mathcal{T}(X^{\pi_b},Y,Z)$ for
$b=1,\dots,B$, where $X^{\pi_b}$ is the source after resampling $\pi_b$, \(Y\)
is the target, and~\(Z\) collects the conditioning variables. The~$p$-value is
the fraction of null values strictly greater than $t$,
\begin{equation}
p=\frac{1}{B}\sum_{b=1}^{B}\mathbf{1}\{\mathcal{T}_b>t\},
\end{equation}
where \(\mathbf{1}\) is the indicator function.
The standardised effect size is
\begin{equation}
t_{\mathrm{score}}=\frac{t-\bar{\mathcal{T}}}{s_{\mathcal{T}}},\qquad
\bar{\mathcal{T}}=\frac{1}{B}\sum_{b=1}^{B}\mathcal{T}_b,\qquad
s_{\mathcal{T}}^2=\frac{1}{B-1}\sum_{b=1}^{B}(\mathcal{T}_b-\bar{\mathcal{T}})^2.
\end{equation}
Percentiles of the null sample give the reported confidence intervals. The~implementation follows the Python reference, so the smallest positive $p$-value
is $1/B$ and $p=0$ means that no surrogate exceeded the observed~value.

The default permutation scheme shuffles the source, which preserves each marginal
distribution but destroys dependence. It therefore targets the null hypothesis of
iid exchangeability~\cite{schreiber2000measuring}. For~autocorrelated series, the~library provides three extensions. Block permutation splits the source into
contiguous blocks of length $L$, shuffles the block order, and~concatenates the
blocks. Circular shift rotates the source by a random offset. Block bootstrap
resamples contiguous blocks with replacement. The~first two preserve short-range
temporal structure and are appropriate for time-series null hypotheses
~\cite{theiler1992testing,schreiber2000surrogate}. Block resampling is described
by~\cite{kunsch1989jackknife,politis1994stationary}. The~block bootstrap in this
library is a block-wise source-resampling null that preserves short-range source
structure. It is not a joint bootstrap and should not be read as a confidence
interval. For~related bootstrap-based uncertainty estimates for time-series
entropy measures, see~\cite{ramirez2021cross}. The~block length should reflect
the autocorrelation time of the~source.

For transfer entropy the effective value subtracts the estimate obtained from a
resampled source~\cite{marschinski2002analysing}. The~default uses a single draw.
The multi-surrogate extension averages over $B$
draws and reports the standard error $s_{\mathcal{T}}/\sqrt{B}$. The~tests are available only
where permuting a source is meaningful. A~$p$-value measures significance and not
effect size, so the significance level and any correction for multiple
comparisons are left to the analyst. Seeded draws are reproducible and independent
of execution order, and~the optional parallel path returns identical~results.

\section{Validation Against Analytical~Solutions}
\label{app:validation}

The continuous estimators were checked against analytically known values. The~test data are multivariate Gaussian with unit variances and a common pairwise
correlation $\rho=0.6$ for entropy and mutual information and $\rho=0.8$ for
conditional mutual information. For~a $d$-dimensional Gaussian with covariance
$\Sigma$ the differential entropy is
\begin{equation}
H(X)=\frac{1}{2}\log\left((2\pi e)^d|\Sigma|\right),
\end{equation}
and the corresponding R\'enyi and Tsallis entropies are
\begin{align}
H_{\alpha}(X)&=\frac{d}{2}\log(2\pi)+\frac{1}{2}\log|\Sigma|
-\frac{d}{2(1-\alpha)}\log\alpha,\\
H_{q}(X)&=\frac{1-(2\pi)^{d(1-q)/2}|\Sigma|^{(1-q)/2}q^{-d/2}}{q-1}.
\end{align}
For the equicorrelation covariance,
$|\Sigma|=(1-\rho)^{d-1}(1+(d-1)\rho)$. The~mutual information between two
$d$-dimensional blocks with correlation $\rho$ is
\begin{equation}
I(X;Y)=-\frac{d}{2}\log(1-\rho^2),
\end{equation}
and the conditional mutual information of a trivariate equicorrelation Gaussian is
\begin{equation}
I(X;Y\mid Z)=-\frac{1}{2}\log\left(1-\rho_{\mathrm{partial}}^2\right),
\qquad
\rho_{\mathrm{partial}}=\frac{\rho}{1+\rho}.
\end{equation}
For transfer entropy, we use the scalar Gaussian VAR(1)
\begin{equation}
X_t=aX_{t-1}+\varepsilon_{x,t},\qquad
Y_t=bY_{t-1}+cX_{t-1}+\varepsilon_{y,t},
\end{equation}
where \(\varepsilon_{x,t}\) and \(\varepsilon_{y,t}\) are the innovations, with~\(a=b=0.5\), \(c=0.6\), and~unit innovation variances. The~analytical value is
\begin{equation}
T_{X\to Y}=\frac{1}{2}\log\left(1+c^2\frac{\operatorname{Var}(X_{t-1}\mid Y_{t-1})}{\sigma_y^2}\right),
\end{equation}
where \(\sigma_y^2\) is the innovation variance of \(Y\) and the conditional
variance follows from the stationary covariance of the VAR.
This gives $T_{X\to Y}=0.1814$ nats. The~R\'enyi and Tsallis truths are evaluated
at $\alpha=q=1.5$ from the Gaussian entropies~above.

\begin{figure}[H]
\includegraphics[width=0.8\textwidth]{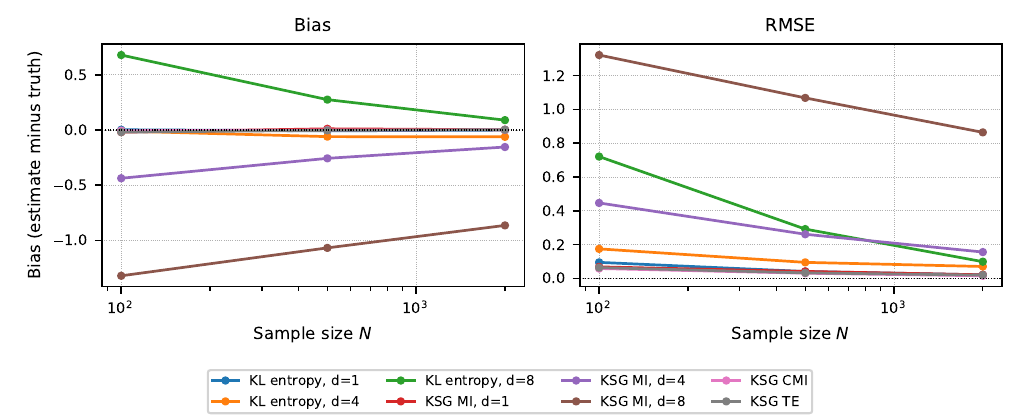}
\caption{Bias
 (\textbf{left}) and RMSE (\textbf{right}) versus sample size for selected estimators.
The low-dimensional estimators are close to the analytical truth at
$N\ge500$. The~high-dimensional estimators remain computationally fast but retain
a finite-sample bias at these sample sizes. \label{fig:validation}}
\end{figure}

The study uses 100 independent replicates for each cell, with~$N=100,500,2000$
samples and $d=1,2,4,8$ for entropy and mutual information. Conditional mutual
information and transfer entropy are scalar. Table~\ref{tab:validation} reports
the mean bias and root-mean-square error, and~Figure~\ref{fig:validation} shows
the dependence on sample size for selected~cells.

\begin{table}[H]
\caption{Validation against analytical solutions. Bias is estimate minus truth
and RMSE is the root-mean-square error, both in nats over 100 replicates.
Entropy and mutual information vary the dimension $d$. Conditional mutual
information and transfer entropy are scalar. \label{tab:validation}}
\centering
\footnotesize

\begin{tabularx}{\textwidth}{lCCCCCCC}
\toprule
& & \multicolumn{2}{c}{\boldmath{$N=100$}} & \multicolumn{2}{c}{\boldmath{$N=500$}} & \multicolumn{2}{c}{\boldmath{$N=2000$}} \\
\cmidrule{3-8}
\textbf{Estimator} & $d$ & \textbf{Bias} & \textbf{RMSE} & \textbf{Bias} & \textbf{RMSE} & \textbf{Bias} & \textbf{RMSE} \\
\midrule
KL & 1 & +0.002 & 0.095 & $-$
0.008 & 0.041 & $-$0.001 & 0.021 \\
KL & 4 & $-$0.010 & 0.175 & $-$0.060 & 0.095 & $-$0.062 & 0.071 \\
KL & 8 & +0.679 & 0.721 & +0.275 & 0.292 & +0.089 & 0.099 \\
R\'enyi & 1 & +0.018 & 0.114 & $-$0.002 & 0.047 & +0.001 & 0.026 \\
R\'enyi & 4 & +0.112 & 0.233 & +0.040 & 0.097 & +0.007 & 0.040 \\
R\'enyi & 8 & +0.806 & 0.855 & +0.447 & 0.461 & +0.275 & 0.280 \\
Tsallis & 1 & +0.008 & 0.058 & $-$0.001 & 0.024 & +0.000 & 0.014 \\
Tsallis & 4 & +0.011 & 0.024 & +0.004 & 0.010 & +0.001 & 0.004 \\
Tsallis & 8 & +0.011 & 0.011 & +0.007 & 0.007 & +0.004 & 0.004 \\
KSG MI & 1 & $-$0.020 & 0.068 & +0.011 & 0.041 & +0.002 & 0.022 \\
KSG MI & 4 & $-$0.437 & 0.447 & $-$0.257 & 0.262 & $-$0.154 & 0.156 \\
KSG MI & 8 & $-$1.320 & 1.322 & $-$1.068 & 1.068 & $-$0.864 & 0.864 \\
Kernel MI & 1 & +0.073 & 0.089 & +0.016 & 0.031 & +0.006 & 0.016 \\
KSG CMI & 1 & $-$0.006 & 0.059 & $-$0.000 & 0.031 & +0.002 & 0.016 \\
KSG TE & 1 & $-$0.019 & 0.064 & $-$0.003 & 0.033 & +0.003 & 0.020 \\
\bottomrule
\end{tabularx}

\end{table}
\unskip

The low-dimensional estimators are essentially unbiased for $N\ge500$. At~$d=8$ the bias is still visible at $N=2000$, most strongly for the KSG mutual
information, which underestimates the analytical value. This is a statistical
property of the estimators and not a runtime issue. The~library's
high-dimensional optimisation is computational. A~stable estimate at high $d$
requires a larger sample, and~the runtime advantage does not remove this
requirement. This is the sense in which the performance claims in
Section~\ref{sec:perf} are~conservative.

\section{Dimensional~Scaling}
\label{app:scaling}

\begin{figure}[H]
\includegraphics[width=\textwidth]{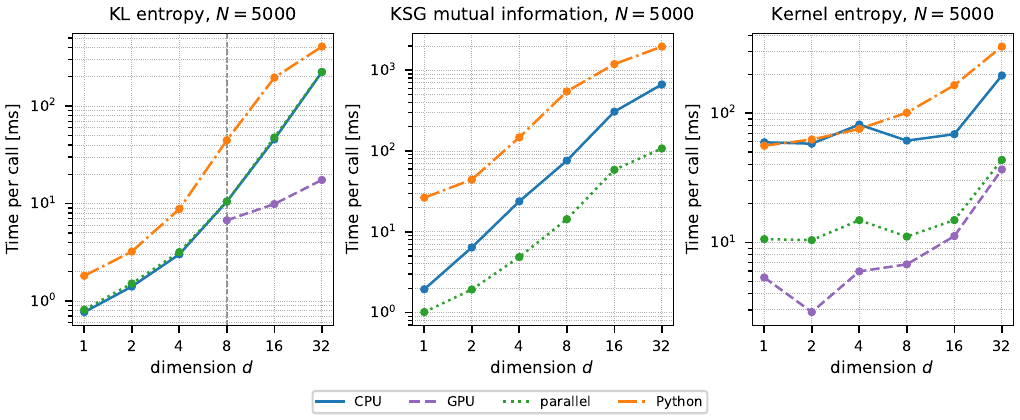}
\caption{KL
 entropy, KSG mutual information and Gaussian-kernel entropy runtime
versus dimension at $N=5000$ for \textit{infomeasure-rs} CPU, GPU and parallel
paths, and~the Python reference. Separate scaling run on an Apple M4 Pro. The~exponential-family GPU path only dispatches at $d\ge8$, marked by the dashed
line, while the kernel gate is a sample-size threshold and is active at every
$d$ shown. The~parallel path coincides with the CPU for KL entropy. KSG is not
covered by the GPU path, so no GPU series is shown.
\label{fig:scaling-dim}}
\end{figure}

This appendix reports a separate scaling run on an Apple M4 Pro. It is not part
of the cross-package benchmark. The~inputs are Gaussian arrays with a fixed seed,
the CPU baseline is single-threaded, and~each cell is the mean of three timed
calls after two warm-up calls. The~dimensions are $d=1,2,4,8,16,32$ and the
sample sizes are $N=1000$ and $N=5000$. Table~\ref{tab:scaling} reports the
$N=5000$, $d=32$ point. Figure~\ref{fig:scaling-dim} shows $N=5000$, the~regime
where the GPU and parallel paths~engage.

\begin{table}[H]
\caption{Separate scaling run on an Apple M4 Pro at $N=5000$ and $d=32$.
Runtime is the mean of three timed calls in milliseconds. Speed-up is relative to
the Python reference. The~GPU entry uses the auto-dispatch build, which only
engages above its dispatch gates. KSG stays on the CPU because the dense GPU
path does not cover KSG. \label{tab:scaling}}
\centering
\small

\begin{tabularx}{\textwidth}{llCC} 
\toprule
\textbf{Measure} & \textbf{Build} & \textbf{Runtime [ms]} & \textbf{Speed-up vs. Python} \\
\midrule
KL entropy & CPU & 224.0 & 1.8 \\
KL entropy & GPU & 17.5 & 23.4 \\
KL entropy & parallel & 224.3 & 1.8 \\
KL entropy & Python & 408.8 & 1.0 \\
KSG MI & CPU & 667.4 & 2.9 \\
KSG MI & GPU & 670.0 & 2.9 \\
KSG MI & parallel & 107.6 & 18.3 \\
KSG MI & Python & 1965.7 & 1.0 \\
Kernel entropy & CPU & 195.5 & 1.7 \\
Kernel entropy & GPU & 36.6 & 8.9 \\
Kernel entropy & parallel & 43.5 & 7.5 \\
Kernel entropy & Python & 327.3 & 1.0 \\
\bottomrule
\end{tabularx}

\end{table}
\unskip

The CPU runtime grows with dimension for the k-d tree measures. For~KL entropy
at $N=5000$ it rises from 0.8\,ms at $d=1$ to 224.0\,ms at $d=32$. The~GPU path
engages from $d=8$ and reduces the $d=32$ runtime to 17.5\,ms. For~KSG mutual
information at $N=5000$ the CPU runtime rises from 1.9\,ms to 667.4\,ms, and~the
parallel path reduces the $d=32$ runtime to 107.6\,ms. The~Python reference rises
from 1.8 to 408.8\,ms for KL entropy and from 26.4 to 1966\,ms for KSG mutual
information over the same range. At~$d=32$ the CPU advantage over Python narrows
for KL entropy, as~the neighbour search degrades in both implementations. The~GPU path remains about 23 times faster than~Python.

The Gaussian-kernel entropy shows the other acceleration path. At~$N=5000$ the
CPU runtime ranges from 58 to 196\,ms over the sweep, the~GPU path from 5.3 to
36.7\,ms, and~the parallel path from 10.6 to 43.5\,ms. The~kernel gate is a
sample-size threshold, so the GPU dispatches at every dimension~here.

The k-d tree therefore remains the CPU path at $d=8$, 16 and 32, but~its
traversal cost grows. The~dense GPU path and the parallel path recover part of
that cost, which is the practical meaning of the dimensional optimisation in
Section~\ref{sec:package}.

\section{Parameters of the Illustrative~Examples}
\label{app:example-params}

Table~\ref{tab:example-params} lists the integration, sampling, and~estimator
parameters used for the three examples in Section~\ref{sec:examples}. All
measure calls use natural logarithms and a noise level of \(10^{-8}\) unless
stated otherwise. In~the table, \(x_t\) is the logistic state and \(r\) the map
parameter; \(\omega_i\) are the Kuramoto natural frequencies and \(K\) the
coupling; \(a\), \(b\) and \(c\) are the R\"ossler parameters; \(\omega_1\)
\(\omega_2\) are the R\"ossler natural frequencies; and \(\varepsilon\) is the
coupling. \(\mathcal N\) denotes the normal~distribution.

\begin{table}[H]
\caption{Parameters
 of the illustrative~examples. \label{tab:example-params}}
\centering
\footnotesize
\setlength{\tabcolsep}{3pt}
\renewcommand{\arraystretch}{1.05}
\renewcommand{\tabularxcolumn}[1]{p{#1}}

\begin{tabularx}{\textwidth}{p{3.4cm}  p{10cm}}

\toprule
\textbf{Parameter} & \textbf{Value} \\
\midrule
\arrayrulecolor{gray!30}
\multicolumn{2}{l}{\textbf{Logistic map}} \\
\hline
Map & \(x_{t+1}=r x_t(1-x_t)\) \\
\hline
Initial condition & \(x_0=0.4\) \\
\hline
Sampling & 3000 samples after 2000 transient~steps \\
\hline
Sweep & \(r=3.5\) to \(4.0\) in steps of~0.002 \\
\hline
Entropy estimators & Order-3 permutation entropy. Gaussian kernel entropy with bandwidth 0.02.  Kozachenko and Leonenko entropy with \(k=4\) as a cross-check, not~plotted. \\
\hline
\multicolumn{2}{@{}l}{\textbf{Kuramoto oscillators}} \\
\hline
Population & \(N=24\), \(\omega_i\sim\mathcal N(1,0.1)\) \\
\hline
Integration & RK4, \(dt=0.01\), 20\,000 steps after 5000 transient~steps \\
\hline
Sweep & \(K=0\) to \(6\) in steps of~0.25 \\
\hline
Estimators & KSG mutual information with \(k=4\). Gaussian kernel mutual information with bandwidth 0.2. KSG transfer entropy with \(k=4\) and lags \(1,1,1\). \\
\hline
\multicolumn{2}{@{}l}{\textbf{Coupled R\"ossler systems}} \\
\hline
Parameters & \(a=0.15\), \(b=0.2\), \(c=10\), \(\omega_1=1.015\), \(\omega_2=0.985\) \\
\hline
Integration & RK4, \(dt=0.05\), 24\,000 steps after 4000 transient steps, stride~4 \\
\hline
Ensemble & Ten trajectories with perturbed initial~conditions \\
\hline
Sweep & \(\varepsilon=0\) to \(0.2\) in steps of~0.01 \\
\hline
Estimators & KSG transfer entropy with \(k=4\). Gaussian kernel transfer entropy with bandwidth 0.2. Ordinal transfer entropy of order 3. All use lags \(1,1,1\). \\
\hline
Parametric sweeps & R\'enyi \(\alpha=0.6\) to \(2.0\) and Tsallis \(q=0.6\) to \(2.0\) in steps of 0.1 on a single 12\,000-step trajectory with \(k=4\). \\
\hline
\arrayrulecolor{black}
\bottomrule
\end{tabularx}
\end{table}

\end{document}